\documentclass{aa}
\usepackage{graphicx}
\usepackage{txfonts}
\usepackage{natbib}

\def\A {\hbox{\textsf A}}
\def\B {\hbox{\textsf B}}
\def\C {\hbox{\textsf C}}
\def\D {\hbox{\textsf D}}
\def\E {\hbox{\textsf E}}
\def\F {\hbox{\textsf F}}
\def\G {\hbox{\textsf G}}
\def\meth {\hbox{CH$_3$OH} }
\def\kms {\hbox{km\,s$^{-1}$}} 
\def\hii {\hbox{\ion{H}{ii}} }

\begin{document}
%
   \title{Discovery of two new methanol masers in NGC\,7538}
   \subtitle{Locating of massive protostars}
   \author{M. R. Pestalozzi \inst{1}
          \and
           V. Minier \inst{2,}\inst{3}
          \and
	  F. Motte \inst{3}
	  \and
          J. E. Conway \inst{4}}

   \offprints{M. Pestalozzi, michele@star.herts.ac.uk}

   \institute{School of Physics, Astronomy and Mathematics, University of
              Hertfordshire, AL10 9BS Hatfield, UK, \\
	      \email{michele@star.herts.ac.uk}
                    \and
              Service d'Astrophysique, DAPNIA/DSM/CEA Saclay,
              91191 Gif-sur-Yvette, France 
                    \and
	      AIM, UMR 7158, CEA-CNRS-Universit\'e Paris
              VII, CEA/Saclay, 91191 Gif sur Yvette, France
		    \and
              Onsala Space Observatory, 439 92 Onsala, Sweden}

   \date{In preparation for submission to A\&A }

\authorrunning{Pestalozzi et al.}
\titlerunning{Discovery of two new methanol masers in NGC\,7538}


\abstract{NGC\,7538 is known to host a
  6.7 and 12.2\,GHz methanol maser cospatial with a Ultra Compact (UC)\,\hii
  region, IRS~1.}{We report on the
  serendipitous discovery of two additional 6.7\,GHz 
  methanol masers in the same region, not associated
  with IRS~1.}
{Interferometry maser positions are compared with recent single-dish and
  interferometry continuum observations.}
{The positions of the masers agree to high
  accuracy with the 1.2\,mm continuum peak 
  emission in NGC\,7538~IRS~9 and NGC\,7538~S. This
  clear association is also confirmed by the positional agreement of the
  masers with existing high resolution continuum observations at cm and/or mm
  wavelengths.}{Making use
  of the established strong relation between methanol masers and high-mas
  star formation, we claim that we have accurately positioned the high-mass
  protostars within the regions where they are detected. The variety of
  objects hosting a 6.7\,GHz methanol maser in NGC\,7538 shows that
  this emission probably traces different evolutionary stages within the
  protostellar phase. }


   \maketitle
%

\section{Introduction}

The NGC\,7538 nebula  is a furnace of intense massive star formation located
at some 2.7~kpc from the Sun. At least 11 infrared sources have been
identified embedded within it (Fig.~\ref{fig:ngc_allIR}, \citealt{ojh04} and
references therein). These IR sources lie in four regions hosting a large
range of different types of young stellar objects (YSOs) with signs of
decreasing age moving from the North-West to the South-East. In the far
North-West the first star formation region contains the IR sources IRS~5, 6, 7
which are associated with a developed 
\hii region of $\sim3$~pc in size. IRS~6 has been proposed as its
exciting source \citep{ojh04}. A further star formation region can be
identified with the IR sources IRS~1-3, where infrared emission  at K-band and
long-ward is dominated by IRS~1 \citep{deb05}. IRS~1 is known to power a
Ultra Compact (UC)\,\hii region, an early stage of massive star
formation. Masers of different species have also been detected toward this
source (OH, H$_2$O, \meth, see \citealt{min98} and  references therein). A
third star-forming region is centered on NGC\,7538~S and also contains 
IRS~11, 10$\arcsec$ to its northwest (Fig. \ref{fig:ngc_allIR}). Coincident
with NGC\,7538~S OH and H$_2$O masers have been detected
\citep{kam90}. Finally, furthest to the South-East, the fourth known region of
star formation contains the deeply embedded IR source IRS~9, believed to be in
a stage prior to the formation of a UC \hii region. H$_2$O masers were also
detected toward this source \citep{dav98,kam90}. 

Methanol masers are indicative of the earliest stages of massive star
  formation. They arise from cold, embedded molecular clumps and in some cases
  from dark infrared clouds \citep{hil05,pur05}. These clumps are often
  interpreted to be protoclusters of high mass YSOs, i.e. precursors of OB
  associations (e.g. \citealt{mot03,min05}). High angular resolution
  observations of methanol maser environments reveal that these masers are
  associated with hot cores, UC\,\hii regions and very often do not
  coincide with radio continuum emission from bright Ultra Compact (UC)\,\hii
  regions (\citealt{wal98,min01}). The detection of a methanol maser in such
  regions is 
  a clear indication of the presence of a deeply embedded hot core that is
  characterised by large methanol abundance and 100-200\,K gas and dust
  temperature. These physical conditions are in agreement with those derived
  from maser modelling, as e.g. \citealt{sob97}. The high astrometric accuracy
  of VLBI observations of methanol masers gives thus the best 
  estimate of the position of the high-mass protostellar core at a scale of
  $\sim$ 10\,AU. One clear example
  of this fact is given by the main component of the \meth{} maser in
  NGC\,7538 (component \A{} in Fig. \ref{fig:spectrum}): the maser pinpoints
  the massive  protostellar core within the IRS~1 submm clump to
  milliarcsecond accuracy 
  (1\,mas at $\sim$3\,kpc is $\sim$3\,AU, see \citealt{pes04a}).

In this letter we report on the serendipitous detection of 6.7\,GHz methanol
masers in the third and fourth star formation regions in NGC\,7538, cospatial
with  NGC\,7538~S and IRS~9. Supported by what exposed above, we argue that
with this discovery we have accurately positioned the two massive protostars
in those regions. This is then confirmed by the coincidence of the new masers
with the peak of the 1.2\,mm dust emission coming from the thick cocoon
surrounding them. 

\section{Observations and data reduction}
\label{sec:data}

The data used for this paper was taken in two runs, with the principal aim of
studying the known 6.7\,GHz methanol maser emission toward NGC\,7538
(i.e. spectral components \A~$\!\!$-\E~$\!$, see Fig. \ref{fig:spectrum}). As
explained below, the results presented here refer to the serendipitous
discovery of two new maser features, unseen until now in any interferometry
experiment.  

The first data set (run1, November 4$^{th}$-5$^{th}$ 2004) was obtained during 
 a joint experiment including the European VLBI Network\footnote{The EVN
  is a joint facility of European, Chinese, South African and others radio
  astronomy institutes} and the three antennas of  the MERLIN\footnote{The
  Multi-Element Radio Linked Interferometer Network is a National Facility run
  by the University of Manchester, UK} array which were then operational at
5cm wavelength, Cambridge (CM), Knockin (KN) and Darnhall (DA). The spectral
resolution was 1.9 and 3.9~kHz/chan (or 0.087 and 0.175~\kms per channel,
respectively) over 512 channels for the joint EVN and the MERLIN only data,
respectively. Only the MERLIN portion of this data set has so far been 
reduced and is presented here. The second data set (run2, December
11$^{th}$-12$^{th}$ 2004) was a 13 hour MERLIN only experiment, 
including 5 antennas: CM, KN, DA, the 25-m Jodrell Bank Mk2 antenna
(JB) and Tabley (TA). The spectral resolution was 0.976~kHz/chan 
(0.044~\kms per channel) over 512 channels.  In both runs the telescopes were
pointed, and the data correlated, at the position of the brightest methanol
maser feature in IRS~1, feature \A{} in Figure \ref{fig:spectrum} (see also
 Table \ref{tab:DE_char}). 

First order calibration and translation into \verb*|fits| format was performed 
with \verb*|tdproc| at Jodrell Bank. Corrections to the primary
amplitude calibration, phase calibration (self calibration on the
brightest emission channel) and image production were performed in AIPS. 
To locate all the maser emission, large 10x10 arcsec spectral cubes were made 
spanning the full observed velocity range. 

In the cross correlation data from run1 (confirmed then by the analysis of the
data from run2) emission was detected at the velocities of the known spectral
features but also unexpectedly in some other spectral channels. The latter was
large non noise-like side-lobe emission, indicating that maser emission at
these velocities came from sources located outside of the map region. Maps
were then iteratively re-centered until the new features, were brought close
to the image centre.  

The large position offsets of the new features from the correlated position
cause large variations of the visibility phase over the integration time, and
hence introduce amplitude errors on the visibilities. These do not affect
feature position determination, but prevent us from making reliable
images. Features \F{} \& \G{} showed maximal phase changes of 0.6 and
1.8$^{rad}$ over the integration time of 7\,s, producing mean amplitude losses
of $\sim$1 and 8\% respectively, on the baseline in Figure \ref{fig:spectrum}.

The positions of the masers were compared to those of 1.2~mm dust peak
emission that were obtained in a MAMBO-2/IRAM-30m map. The image was obtained
in January 2004 using the MAMBO-2 bolometer array \citep{kre98} installed at
the IRAM 30~m telescope in the on--the--fly mode.  The resulting angular
resolution is {\em HPBW} $\sim 11\arcsec$  and the absolute pointing is
accurate to within $\sim 3\arcsec$.

\begin{figure}
\begin{center}
\caption{General view of the NGC\,7538 region. Left: The underlying colour
  image is a RGB image of J, H, K NIR bands made from the 2MASS archive. The
  grey contours (2, 5, 10, 15, 20, 30, 40, 50 ,60, 70 ,80, 90 \% of peak
  emission, $\sim$4 Jy beam$^{-1}$) show 1.2\,mm dust continuum mapped with
  MAMBO-2/IRAM-30m. White 
  circles indicate the location of the identified infrared sources. White
  crosses indicate the 6.7~GHz \meth{} maser
  positions. Right: The underlying image is the 1.2 mm continuum
  map. 1\,arcmin is 0.78\,pc.}
\label{fig:ngc_allIR}
\end{center}
\end{figure}

\section{Results}
\label{sec:results}

\subsection{\meth{} masers features \F{} \& \G}
\label{sec:res1}

In the following, we denote the two new spectral features found during data
  reduction as \F{} and 
  \G\footnote{The nomenclature of all features 
  follows the one adopted in \citet{min00a}, based on the spatial
  distribution of the masers}.  Their positions and other properties are
summarised in Table \ref{tab:DE_char}. 

\begin{figure}
\begin{center}
\includegraphics[width=8cm]{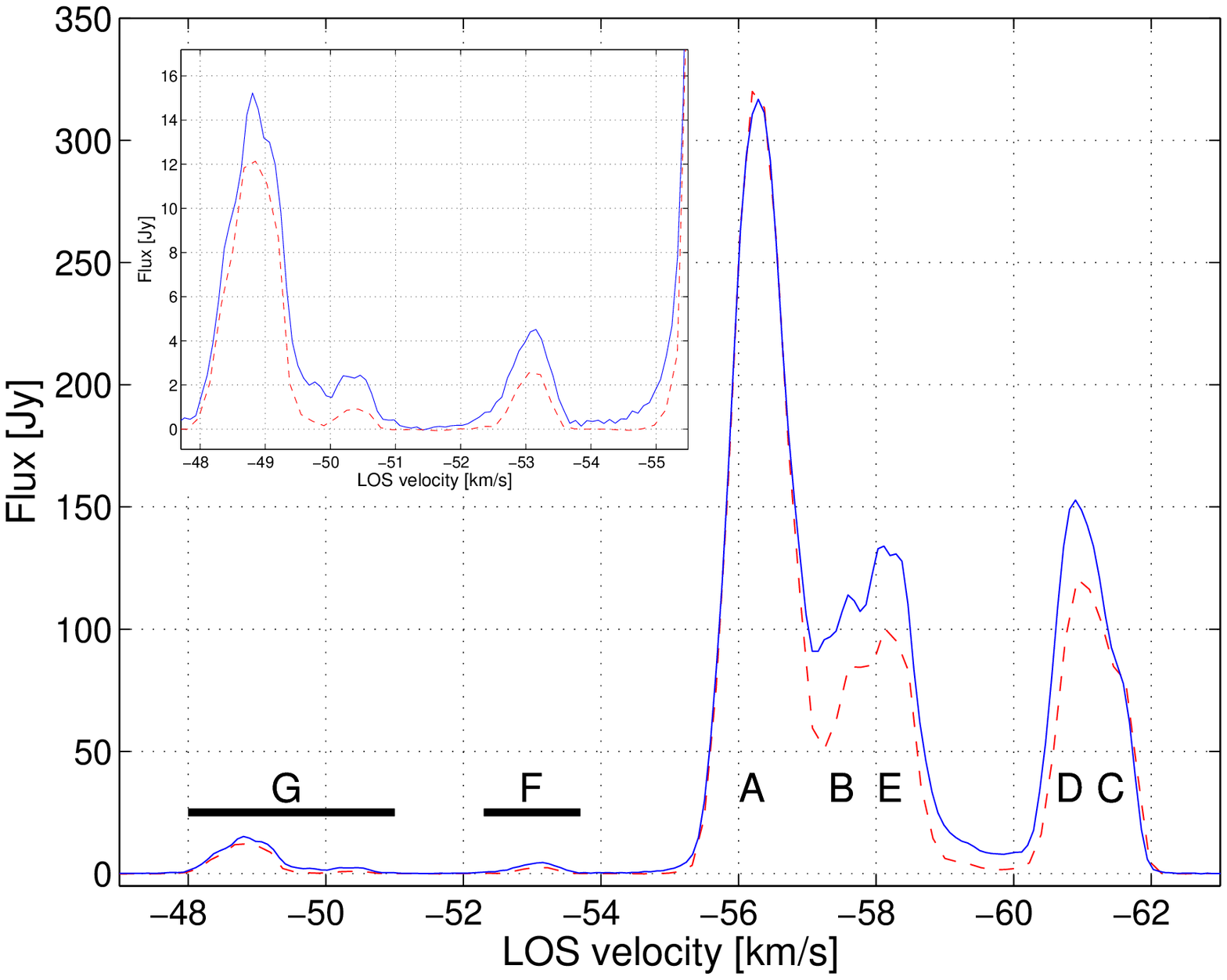}
\caption{Cross correlated spectrum on the KN-DA baseline averaged over the
  whole observing run (dashed line) and
  autocorrelation spectrum from the DA antenna (solid line), both from
  run1. Minimal fringe spacing on the KN-DA baseline is $\sim$190~mas. Features
  \A,\B,\C,\D and \E are located within 0.5 arcsec of IRS~1. The inset is a
  zoom-in of the velocity region including features \F{} \& \G. The
  multiplicity of \F{} \& \G{} is discussed in the text.}
\label{fig:spectrum}
\end{center}
\end{figure}

The comparison of the cross correlated spectrum on the shortest baseline with
the integrated spectrum of the maser emission shows discrepancy in
the peak flux of feature \F{} \& \G{} of $\sim$20\%
(Fig. \ref{fig:spectrum}). Unfortunately, the amplitude errors mentioned in
Section \ref{sec:data} and the unreliable baseline subtraction in the
integrated spectrum introduce errors in the amplitude that can be estimated to
15\%. These prevent us to invoque resolution as responsible for the missing
flux. Our best estimate for the size of \F{} \& \G{} (upper limit) is about
half the fringe spacing of the shortest baseline, i.e. $\sim$85\,mas, or
260\,AU, which is to be taken as an upper limit. Note that in the higher
spectral resolution data of run2 (not shown here), \F{} \& \G{} show clear
multiplicity (at least two and three components for F{}\&
\G{} respectively). Observations pointed at the new masers are necessary to
obtain reliable and accurate maps of all features. 

A review of archival EVN VLBI  autocorrelation data for all antennas 
except Effelsberg shows that both \F{} and \G{} are clearly present in 
all experiments since 1997. These features always have similar 
flux ratios to spectral feature \A (within a variation of the latter of some
10\%). This indicates that the emission from \F{} 
and \G{} is probably non-variable. The new features are   
not detected in the autocorrelations from the  Effelsberg 100\,m antenna
because they fall outside the HPBW of the primary beam at 5cm
(i.e. 1.6\,arcmin). Note that despite the reduction in amplitude due
to the primary beam attenuation a hint of  feature \G{} was present in the
discovery spectrum for NGC\,7538 taken with the 140\,foot Green 
Bank telescope \citep{men91a}. 

\subsection{\meth masers in NGC\,7538~IRS~9 \& S}
\label{sec:ass}

Figure \ref{fig:ngc_allIR} shows a general view of the NGC\,7538 region. 
The \meth{} masers are clearly associated with the most embedded sites of
high-mass star formation: IRS~1, IRS~9 and S.  Masers in IRS~1 coincide with
the 1.2\,mm continuum peak emission (within the position accuracy) tracing 
cold dust emission from massive and deeply embedded protoclusters.

Maser feature \F{} coincides with the 1.2\,mm dust continuum emission peak
within $\sim\!2\arcsec$ (Fig. \ref{fig:ngc_allIR}). This is less than the
position accuracy of the 1.2\,mm 
data ($\sim\!3\arcsec$), and therefore we can state that they are
cospatial. The methanol maser also coincides with other signposts of high-mass
star formation in NGC\,7538~S as reported in the literature, within the
position accuracies 
\citep{san04,san03,kam90}. In \citet{san03} this protocluster/cloud is stated
to be one of the most massive of the region, with 1000\,M$_{\odot}$ estimated
within a circle of 20\arcsec (or 50000\,AU) in diameter. The
authors 
of that paper also report of a large rotating torus of some 14000\,AU in
radius. Single grey--body Spectral Energy Distribution (SED) fitting of the
cold dust emission in the FIR/(sub)mm/mm emission with an aperture of
60$\arcsec$ of this source gives a dust temperature of $\sim$35\,K and an
emissivity index $\beta=1.6$. This source lies some 10$\arcsec$ to the south
of IRS~11, where a CO outflow is inferred by the distribution of the magnetic
field orientation \citep{kam91}. On the basis of the above multiple evidence,
we consider the position of the methanol maser to be the most reliable for the
protostellar object in NGC\,7538~S. 

Maser feature \G{} is coincident within 1$\arcsec$ with the peak of the
1.2\,mm dust emission. The position of \G{} also agrees, within the
position inaccuracies, with the high resolution positions of the peak emission
of the continuum at 4.8 to 107\,GHz \citep{vdt00,vdt05}. Furthermore, one
water vapour maser spot as well as one class I methanol maser spot seem also to
coincide, within the positional accuracies, with feature \G{} 
\citep{kam90,san05}, although with an offset in velocity of some
6\,\kms. Because of the fact that class I 
methanol masers are mainly found in outflows \citep{kur04}, we argue that the
newly detected 
maser \G{} marks the driving source of the outflow (together
with the continuum), while the other masers arise in the outflow
shocks. This is supported by the geometry proposed in \citet{san05} for that
source: we might observe the outflow from IRS~9 pole-on. The protostellar
object in IRS~9 is accurately located both with the class II methanol maser
and the cm/mm continuum.

\section{Discussion}

The presence of three methanol maser sources associated to three different
objects in the same star formation region offers a unique
opportunity to study the evolutionary details of methanol maser bearing
sources. 

The methanol maser phase is considered to be relatively short as
compared to the lifetime of the massive protostar, as short as a few
10$^{4}$\,yr \citep{vdw05}. This implies that the sources hosting methanol
masers at 6.7\,GHz should show very similar properties. The methanol masers in
NGC\,7538 (features \A, \G{} \& 
\F) on the other hand seem to mark massive young stellar objects at different
stages of evolution: an IR-quiet protostellar object (NGC\,7538~S), an
IR-bright protostellar object (IRS~9) and an UC\hii region (IRS~1). The same
distinction seems to be shown by the flux densities of the masers: 300, 15,
5\,Jy for \A, \G, \F{} respectively.

We propose 2 scenarios for this apparent sequence. The first and most
intuitive is the evolutionary sequence. Feature \F{} might mark the
youngest object in the region, feature \A{} the most evolved but still within
the protostellar phase. This scenario is also supported by the presence of
discs/outflows in the maser sources: in \F{} a large scale rotating torus is
detected but no outflow \citep{san04}; in \G{} a powerful, probably very
young, outflow is mapped \citep{san05}; in \A{} a clear disc/outflow geometry
is visible (e.g. \citealt{gau95,pes04a,deb05}). More evidence has to be
collected on the capacity of methanol masers to provide such accuracy in the
age the sources hosting them. 

The second scenario invokes geometry. The difference in 1.2\,mm dust emission
peak flux among the methanol maser bearing sources could be due to different
viewing angles to the source. IRS~1 is seen edge-on, showing the thick dust in
its disc \citep{gau95,pes04a}. IRS~9 is suggested to be seen close to pole-on
\citep{san05}: the disc would be seen almost face-on, not contributing
significantly to the 1.2\,mm flux. NGC\,7538~S is still deeply embedded, so
dust is close to uniformly distributed around the source. This interpretation
seems to be more difficult to accept, because e.g. IRS~1 suffers of
significant contamination by the very close IRS~2 and IRS~3.

\begin{table}
\caption{Position, line of sight (LOS) velocity ($v$) and offset ($r$)
  from feature \A{} of the two maser components  \F{} and \G. The
  uncertainties in the positions and in the velocity widths are 50\,mas
  (135\,AU) and 0.044\,km s$^{-1}$ respectively.} 
\begin{tabular}{@{}cr@{.}lr@{.}lcc@{}}
\hline \hline
Masers      & \multicolumn{2}{c}{RA}        &  \multicolumn{2}{c}{DEC}    &
$v$ / $FWHM$          & $r$  \\ 
   &   \multicolumn{2}{c}{J2000} & \multicolumn{2}{c}{J2000} & km s$^{-1}$ &
arcmin / pc  \\ 
\hline
\A & 23 13 45&363 & 61 28 10&55 & --56.2 / 1.0 & \\
\F & 23 13 44&86 & 61 26 48&1 & --53.2 / 0.5 & 1.36 / 1.06 \\
\G & 23 14 01&78 & 61 27 19&7 & --48.8 / 0.9 & 2.14 / 1.67 \\ 
\hline
\end{tabular}
\label{tab:DE_char}
\end{table}


\section{Summary and Conclusions}
\label{sec:concl}

From the high spatial resolution MERLIN observations of 6.7\,GHz methanol maser
emission toward the massive star forming region NGC\,7538 we have detected and
positioned two new spectral features. We note that this achievement
was possible only thanks to the intermediate resolution of the MERLIN
with its shortest baselines. We conclude the following:  
\begin{itemize}
\item on the basis of the accepted idea that methanol masers are the most
  efficient tracers of young and embedded 
  massive protostars, the accurate location (to within 50\,mas or 135\,AU) of
  \F{} and \G{} is 
  equivalent with the accurate location of the massive protostars in
  NGC\,7538~IRS~9 and NGC\,7538~S;
\item the variety of objects marked by methanol maser emission could be due to
  different ages or viewing angle geometry. To test these two scenarios, more
  accurate observations (targeted toward \F{} \& \G) are required. 

\end{itemize}


\begin{acknowledgements}
M.P. thanks Tom Muxlow for his help during the data calibration at
the Jodrell Bank Observatory. We thank all the participants of the NGC\,7538
collaboration for the very fruitful discussions that contributed to the
acheivement of this paper.
\end{acknowledgements}

\bibliography{methanol,varia,othermasers,starformation+IR,surveys_cat,tech} 
\bibliographystyle{aa}

\end{document}